\documentclass[12pt]{article}
\usepackage{amssymb}
\usepackage{amsmath}
\usepackage{graphicx}
\newcommand{\Fc}{\mathcal{F}}
\newcommand{\ead}[1]{\vspace*{5pt} {E-mail: \tt{#1}}}

\begin{document}

\title{\Large\textbf{EXACT SOLUTIONS FOR NONLOCAL NONLINEAR
FIELD EQUATIONS IN COSMOLOGY}}

\author{\textbf{S.~Yu.~Vernov}\\
Skobeltsyn Institute of Nuclear Physics, Moscow State University,\\
Leninskie Gory, GSP-1, 119991, Moscow, Russia\\
\small\ead{svernov@theory.sinp.msu.ru}}

\date{}
\maketitle

\begin{abstract}
A method for the search of exact solutions for equation of a nonlocal
scalar field in a non-flat metric is considered. In the
Friedmann--Robertson--Walker metric the proposed method can be used in
the case of an arbitrary potential, with the exception of linear and
quadratic potentials, and allows to get in quadratures solutions, which
depend on two arbitrary parameters. Exact solutions have been found for
an arbitrary cubic potential, which consideration is motivated by the
string field theory, as well as for exponential, logarithmic and power
potentials. It has been shown that one can add the $k$-essence field to
the model to get exact solutions for all Einstein equations.
\end{abstract}

\section{GRAVITATIONAL MODELS WITH A NONLOCAL SCALAR FIELD}

Nonlocal cosmological models, based on the string field theory
(SFT)~\cite{SFT-review,IA_TMPh} and $p$-adic string theory~\cite{padic} are actively
developed~\cite{IA1}--\cite{GK}. Characteristic properties of
nonlocal models are the null energy condition violation and arising of
phantom fields, which is connected with high derivative  terms.
 Local models with phantom fields are considered as
physically unacceptable ones because of a problem
quantum instability~\cite{AV-NEC,KKLM}.
In~papers~\cite{SW,CreminelliDE} the instability problem is reduced to such choosing of effective
theory parameters that the instability turns out to be essential
only at times that are not described in the framework of the effective
theory approximation. It is mathematically expressed in that the terms, which result to instability, are treated as corrections essential only at small
energies below the physical cutoff. The given approach allows considering
such effective theories as physically acceptable with the presumption
that an effective theory admits immersion in a fundamental theory, for example, the string field theory.

The interest in cosmological models related to open string field
theory~\cite{IA1} is provoked by the possibility of obtaining solutions
describing transitions from a perturbed vacuum to the true vacuum (the
so-called \textit{rolling solutions}).   After all massive fields
(or some of the lower massive fields) are integrated out using the
equations of motion, the open string tachyon gets a potential with a
nontrivial vacuum corresponding to the minimal value of energy.
In the dark energy model~\cite{IA1} (see also~\cite{AKV,AJ,AKV2,VernovTMF}) it is
implied that the Universe is a slowly decomposing $D3$--brane, whose
decay is described by an open string tachyon mode.  According to the
Sen conjugation~\cite{SFT-review}, the tachyon motion from an unstable
vacuum to the stable vacuum describes the $D$-brane transition to the
true vacuum.  In fact one obtains a nonlocal potential with a nonlocality
parameter determined by the string scale. Using a suitable redefinition
of the fields, one can make the potential local, at that the kinetic
term becomes nonlocal. This nonstandard kinetic term leads to a
behavior similar to the behavior of a phantom field, and
it can be approximated with a phantom kinetic term. Hence, the behavior
of an open string tachyon can be effectively simulated by a scalar
field with a negative kinetic term~\cite{AJK}.

The string field theory also gives asymptotic conditions on the
solutions~\cite{AKV,AJ,AKV2,VernovTMF}.  Special interest is represented with solutions,
which are bounded on the whole of real axis and have a nonzero asymptotic
at $t\to +\infty$. In this paper such solutions of the field equation
in the Friedmann--Robertson--Walker  metric have been found for
cubic and logarithmic potentials.

Let us consider the gravitational model with a nonlocal scalar field
$\phi$, which is described by the following action:
\begin{equation}
S=\int d^4x\sqrt{-g}\alpha^{\prime}\left(\frac{R}{16\pi
G_{\mathrm N}}+\frac1{ g_\text{o}^2}\left(\frac12\phi
\Fc(\Box_g)\phi-V(\phi)\right)-\Lambda\right), \label{action_model}
\end{equation}
where  $G_{\mathrm N}$ is the Newtonian gravitational constant
($8\pi G_{\mathrm N}=1/M_{\mathrm P}^2$,
$M_{\mathrm P}$ is the Planck mass), $\alpha^{\prime}$ is the
string length squared, $g_\text{o}$ is the open string coupling constant. We
use the signature
$(-,+,+,+)$, the matrix $g_{\mu\nu}$
is the metric tensor, $g$ is the determinant of
$g_{\mu\nu}$, $R$ is the scalar curvature.
The potential $V(\phi)$  is a twice
continuously differentiable function, the cosmological constant is considered as a part of the potential
$V(\phi)$. The d'Alembert  operator
$\Box_g$ is applied to scalar functions and can be written as follows:
\begin{equation}
\Box_g  = \frac{1}{\sqrt{-g}} \partial_\mu
\sqrt{-g}g^{\mu\nu}\partial_\nu \,.
\end{equation}

The scalar field $\phi$ is dimensionless, while $[\alpha^{\prime}]=\mbox{cm}^2$ and $[g_\text{o}]=\mbox{cm}$. It is convenient to introduce dimensionless
coordinates $\bar{x}_\mu=x_\mu/\sqrt{\alpha^{\prime}}$,
the dimensionless gravitational constant $\overline{G}_{\mathrm N}=G_{\mathrm N}/\alpha^{\prime}$, and
the dimensionless coupling
constant $\bar g_\text{o}=g_\text{o}/\sqrt{\alpha^{\prime}}$. The curvature scalar calculated in dimensionless coordinates is denoted as $\bar{R}$, the  corresponding d'Alembert operator is marked as ${\overline{\Box}}_g$. We get action~\eqref{action_model} in the following form:
\begin{equation*}
S=\int d^4 \bar{x} \,\sqrt{-g}\biggl(\frac{\bar{R}}{16\pi
{\overline G}_{\mathrm N}}+\frac{1}{{\bar g}_\text{o}^2}\biggl(\frac{1}{2}\phi\,\Fc
({\overline\Box}_g)\phi
-V(\phi) \biggr)\biggr),
\end{equation*}
In the following formulae we always use dimensionless
coordinates and parameters and omit bars over them.

The function $\Fc$ is assumed to be an analytic function on whole complex plane (\textit{i.e.} an entire function), therefore,
one can represent it by the convergent series expansion:
\begin{equation}
\Fc(\Box_g)=\sum\limits_{n=0}^{\infty}f_n\Box_g^{\;n}.
\end{equation}

From action (\ref{action_model}) we obtain the following equations
\begin{eqnarray}
&&G_{\mu\nu}=8\pi G_{\mathrm N}T_{\mu\nu}-8\pi G_{\mathrm N}\Lambda g_{\mu\nu},
\label{EOJ_g}\\
&&\Fc(\Box_g)\phi=\frac{dV}{d\phi}, \label{Equphi}
\end{eqnarray}
where $G_{\mu\nu}$ is the Einstein tensor.

The energy--momentum (stress) tensor $T_{\mu\nu}$ is calculated
by the standard formula
\begin{equation}
\label{TEV} T_{\mu\nu}={}-\frac{2}{\sqrt{-g}}\frac{\delta{S}}{\delta
g^{\mu\nu}}=\frac{1}{ g_\text{o}^2}\Bigl(E_{\mu\nu}+E_{\nu\mu}-g_{\mu\nu}\left(g^{\rho\sigma}
E_{\rho\sigma}+W\right)\Bigr),
\end{equation}
where
\begin{equation}
E_{\mu\nu}\equiv\frac{1}{2}\sum_{n=1}^\infty
f_n\sum_{l=0}^{n-1}\partial_\mu\Box_g^l\phi\partial_\nu\Box_g^{n-1-l}\phi,
\qquad
W\equiv\frac{1}{2}\sum_{n=2}^\infty
f_n\sum_{l=1}^{n-1}\Box_g^l\phi\Box_g^{n-l}\phi-\frac{f_0}{2}\phi^2+V(\phi).
\end{equation}

The case of a quadratic potential has been studied
in papers~\cite{AJV0711,MN,KV,Vernov2010}. In paper~\cite{AJV0711} the
algorithm of localization of the Einstein equations has been proposed
and exact solutions in the spatially flat Friedmann--Robertson--Walker metric have been found.
This algorithm can be used only in the cases of linear and quadratic
potentials. At the same time from the string theory one gets cubic and the fourth degree potentials, so, a quadratic potential can be considered
only as an approximation~\cite{KV}. In this connection
the search of solutions for equation~(\ref{Equphi}) with a cubic potential in the Friedmann--Robertson--Walker metric
is actively conducted~\cite{Joukolskaya0902,CN}.

 In this paper we propose
the method of the search of solutions for (\ref{Equphi}) in the Friedmann--Robertson--Walker
metric, which gives exact two-parameter solutions in the close
form or in quadratures. This method allows to find solutions for an arbitrary
potential $V(\phi)$, with the exception of the cases of linear and quadratic
potentials. In the paper we consider the case of cubic potential, which
is connected with the string theory  (Section~3), as well as cases of logarithmic,
exponential and power potentials (Section~4).

 Note that in distinguish to the
localization method~\cite{Vernov2010}, which allows to localize all
Einstein equations, this method solves only equation (\ref{Equphi}),
whereas  equations~(\ref{EOJ_g}), after the substitution of the obtained solution, are inconsistent  in the general case.
 In Section~5 we show that the supplement of a scalar $k$-essence field gives an
exact solution of the system of all nonlocal Einstein equations.
The question about possible types of the additional matter and ability to obtain an exact solution for all Einstein equations
in modified gravitation  models, for example, in $f(R)$ gravitational model, without additional matter
requires distinct investigations.

 For cubic and exponential potentials approximate solutions for equation
(\ref{Equphi}) with the string field theory inspired form of
$\Fc(\Box)$ have been found by G.~Calcagni and G.~Nardelli~\cite{CN} as
a generalization of their solutions in the Minkowski
space~\cite{CN_Minkowski}. In distinguish from~\cite{CN} in this paper
we obtain the exact solutions for equation (\ref{Equphi}), in addition,
the Hubble parameter $H(t)$ is a solution of equation, it is not given
a priori.

\section{SOLUTIONS FOR EQUATIONS OF MOTION}

Let us consider nonlocal Klein--Gordon equation in the case of an
arbitrary potential:
\begin{equation}
\Fc(\Box_g)\phi=V'(\phi), \label{equagen}
\end{equation}
where a prime denotes the derivative with respect to $\phi$. A particular
solution of (\ref{equagen}) is a solution of the following system of
local equations:
\begin{equation}
\label{eomN}
\sum\limits_{n=0}^{N-1}f_n\Box_g^{\:n}\phi=V'(\phi)-C,\qquad
f_N\Box_g^{\:N}\phi=C,
\end{equation}
where $N-1$ is a natural number,  $C$ is an arbitrary constant.

In the case $f_1\neq 0$ we can choose $N=2$. In the spatially flat Friedmann--Robertson--Walker
metric with the interval:
\begin{equation} \label{FRW}
{ds}^{2}={}-{dt}^2+a^2(t)\left(dx_1^2+dx_2^2+dx_3^2\right),
\end{equation}
where $a(t)$ is the scale factor, we obtain from (\ref{eomN}) the
following system:
\begin{equation}
\label{eomFr}
f_1\Box_g\phi={}-f_1\left(\ddot\phi+3H\dot\phi\right)=V'(\phi)-f_0\phi-C,\qquad
f_2\Box_g^{\:2}\phi=C.
\end{equation}
The function $H(t)$ is the Hubble parameter: $H\equiv\dot a(t)/a(t)$,
and a dot denotes the time derivative. It is easy to see that at $f_2=0$
solutions can exist only for $C=0$. Let us consider the case $f_2\neq
0$. The first equation of (\ref{eomFr}) can be rewritten in the
following form:
\begin{equation} \label{Hphi}
    H={}-\frac{1}{3\dot\phi}\left(\ddot\phi+\tilde{V}'(\phi)-\frac{C}{f_1}\right),
\end{equation}
where
\begin{equation}
\tilde{V}'(\phi)\equiv \frac{1}{f_1}\left(V'(\phi)-f_0\phi\right).
\end{equation}

Equation
\begin{equation}
(\partial_t^2+3H\partial_t)\left(\ddot\phi+3H\dot\phi\right)=\frac{C}{f_2},
\end{equation}
transfigures to the following form
\begin{equation}
(\partial_t^2+3H\partial_t)\tilde{V}'=
\tilde{V}'''{\dot\phi}^2+\tilde{V}''(\ddot\phi+3H\dot\phi)={}-\frac{C}{f_2}.
\end{equation}

We eliminate $H$ and obtain
\begin{equation}
\dot\phi^2=\frac{1}{\tilde{V}'''}\left(\tilde{V}''\tilde{V}'-\frac{C}{f_1}\tilde{V}''-\frac{C}{f_2}\right).
\label{Localeomgenpot}
\end{equation}
Equation (\ref{Localeomgenpot}) can be solved in quadratures. The obtained
solution depends on two arbitrary parameters $C$ and $t_0$, the latter papameter
corresponds to the time shift.

At $C=0$ we get the following equation
\begin{equation}
\dot\phi^2=\frac{\tilde{V}'\tilde{V}''}{\tilde{V}'''}\equiv
\frac{(V'-f_0\phi)(V''-f_0)}{f_1V'''}, \label{Localeomgenpot0}
\end{equation}
using which one can get solutions at $f_2=0$ as well.

Remark that, the proposed method can not be used in the
cases of linear and quadratic potentials, since for them $\widetilde V'''\equiv 0$.
 Consequently this way of the search of solutions is suited
only for nonlinear in $\phi$ equation (\ref{equagen}). In the case of
linear in $\phi$ equation (\ref{equagen}) the search of solutions is
possible due to the localization method~\cite{AJV0711,MN,Vernov2010}.

\section{CUBIC POTENTIAL}

The case of cubic potential is actively studied, since it is connected
with the bosonic string field theory~\cite{Joukolskaya0902,CN}. Let us
find solutions for equation (\ref{equagen}) at
\begin{equation}
V(\phi)=B_3\phi^3+B_2\phi^2+B_1\phi+B_0,
\end{equation}
where $B_0$, $B_1$, $B_2$, and $B_3$ are arbitrary constants, but
$B_3\neq 0$. We get (\ref{Localeomgenpot}) in the
following form
\begin{equation}
\dot\phi^2=4C_3\phi^3+6C_2\phi^2+4C_1\phi+C_0, \label{LocSystemgenC}
\end{equation}
where
\begin{equation}
C_0=\frac{(B_1-C)(2B_2-f_0)}{6f_1B_3}-\frac{Cf_1^2}{6f_1f_2B_3} ,\qquad
C_2=\frac{2B_2-f_0}{4f_1},
\end{equation}
\begin{equation}
C_1=\frac{6B_3(B_1-C)+(2B_2-f_0)^2}{24f_1B_3},\qquad
C_3=\frac{3B_3}{4f_1}.
\end{equation}

Note, that $C_3\neq 0$ since $B_3\neq 0$. Constants $B_2$ and $f_0$ appear in equation~(\ref{LocSystemgenC}) only in the combination
$2B_2-f_0$.  Using the transformation
\begin{equation}
    \phi=\frac{1}{2C_3}(2\xi-C_2),
\end{equation}
we get the following equation
\begin{equation}
\dot\xi^2=4\xi^3-g_2\xi-g_3, \label{equxi}
\end{equation}
where
\begin{equation}
g_2=3C_2^2-4C_1C_3=\frac{(2B_2-f_0)^2-12B_3(B_1-C)}{16f_1^2},
\end{equation}
\begin{equation}
g_3=2C_1C_2C_3-C_2^3-C_0C_3^2={}-\frac{3B_3C}{32f_2f_1}\,.
\end{equation}
A solution of equation (\ref{equxi}) is either the Weierstrass elliptic
function
\begin{equation}
\xi(t)=\wp(t-t_0,g_2,g_3),
\end{equation}
where $t_0$ is an arbitrary number, or a degenerate elliptic function.
As known~\cite{BE}, the Weierstrass elliptic function is a double
periodic  meromorphic function, which has one double pole in the
fundamental parallelogram of periods. If $\phi(t)$ is an elliptic
function, then the Hubble parameter $H(t)$ is an elliptic function as well.

Let us consider degenerated cases.  At $g_2=0$ and $g_3=0$ the general
solution for equation (\ref{equxi}) is
\begin{equation}
\xi=\frac{1}{(t-t_0)^2},
\end{equation}
therefore,
\begin{equation}
\label{elemsol1}
    \phi_1=\frac{1}{C_3(t-t_0)^2}-\frac{C_2}{2C_3}=\frac{4f_1}{3B_3(t-t_0)^2}-\frac{2B_2-f_0}{6B_3}.
\end{equation}
Substituting $\phi_1$ into (\ref{Hphi}), we get
\begin{equation}
H_1=\frac{5}{3(t-t_0)}. \label{H3}
\end{equation}

From conditions $g_2=0$ and $g_3=0$ it follows that
\begin{equation}
C=0,\qquad
B_1=\frac{(2B_2-f_0)^2}{12B_3}.
\label{24}
\end{equation}

Solutions, which are bounded on the whole real axis and tends to a
finite limit at
 $t\rightarrow \infty$, attract the special interest.  Such a solution is
the following function
\begin{equation}
\phi_2=D_2\tanh(\beta (t-t_0))^2+D_0,
\end{equation}
\begin{equation}
D_2 =\frac{4}{3B_3}f_1\beta^2, \qquad
D_0=\frac{1}{18B_3}\left(3(f_0-2B_2)-16f_1\beta^2\right),
\end{equation}
where $\beta$ is a root of the following equation
\begin{equation}
1024f_2f_1\beta^6+576f_1^2\beta^4+324B_3B_1-27(2B_2-f_0)^2=0.
\label{equbeta}
\end{equation}

Bounded real solutions for equation (\ref{LocSystemgenC}) correspond to
real roots of equation (\ref{equbeta}). Pure imaginary roots of equation
(\ref{equbeta}) correspond to unbounded real solutions for equation
(\ref{LocSystemgenC}), because  $\tanh(\beta t)^2=-\tan(\mathrm{i}\beta
t)^2$. The solution $\phi_2$ exists at
\begin{equation}
C=\frac{1}{36B_3}\left(64f_1^2\beta^4-3(2B_2-f_0)^2+36B_3B_1\right).
\end{equation}
The Hubble parameter has the following form:
\begin{equation*}
H_{2}=\frac{\beta(2\cosh(\beta t)^2-3)}{3\cosh(\beta t)\sinh(\beta
t)}-{}
\end{equation*}
\begin{equation*}
{}-\frac{3B_3(D_2\tanh(\beta t)^2+D_0)^2+(2B_2-f_0)(D_2\tanh( \beta
t)^2+D_0)+B_1}{6f_1D_2\beta \tanh(\beta t)(1-\tanh(\beta t)^2)}.
\end{equation*}

 The parameter $t_0$ is an arbitrary complex number, so, using the equality $\tanh(t+i\pi/2)=\coth(t)$,
one gets the following real solutions
\begin{equation}
\tilde{\phi}_2=D_2\coth(\beta (t-t_0))^2+D_0.
\end{equation}

Note that solutions in terms of hyperbolic functions exist only at
$C\neq 0$, since $g_3=0$ at $C=0$, and solution (\ref{elemsol1}) is the
unique solution in terms of elementary functions.

\section{EXACT SOLUTIONS FOR OTHER TYPES OF POTENTIAL}

\subsection{Logarithmic Potential}
Note that one can get equation (\ref{Localeomgenpot}) in the form
(\ref{LocSystemgenC}), staring from a nonpolynomial potential as well.
Indeed, let
\begin{equation}
V(\phi)=C_1\ln(\alpha\phi),
\end{equation}
where $C_1$ and $\alpha$ are arbitrary constants. The parameter
$\alpha$ does not enter into equation (\ref{Localeomgenpot}):
\begin{equation}
\dot\phi^2=\frac{f_0^2}{2f_1C_1}\phi^4+\frac{C(f_2f_0-f_1^2)}{2f_1f_2C_1}\phi^3+\frac{C}{2f_1}\phi
-\frac{C_1}{2f_1}\,. \label{Equln}
\end{equation}
At $f_0\neq 0$ in the general case the Jacobi elliptic functions are
solutions for (\ref{Equln}). At $f_0=0$ and $C\neq 0$
equation (\ref{Equln}) is a particular case of equation
(\ref{LocSystemgenC}), which solutions are the Weierstrass elliptic
functions.

Let us analyse real solutions in terms of the elementary functions.
Such solutions have been found only at $f_0=0$.

At $C=0$ one gets the following solutions
\begin{equation}
\phi_0(t)={}-\frac{\sqrt{-2C_1f_1}}{2f_1}(t-t_0),\qquad
H_0=\frac{2}{3(t-t_0)}+\frac{f_0}{3f_1}(t-t_0).
\end{equation}

At $C\neq 0$ the following solution exists
\begin{equation}
\phi_{ln}=\tilde{D}_2\tanh^2(A(t-t_0))+\tilde{D}_0,
\end{equation}
where $A$ is an arbitrary number,
\begin{equation*}
\tilde{D}_2={}-\frac{C_1(32f_2A^2-9f_1)}{18Af_1(16f_2A^2-3f_1)\varsigma},
\quad\tilde{D}_0= \frac{C_1}{12Af_1\varsigma}, \quad \varsigma=\pm
\sqrt{\frac{C1}{144f_2A^2-27f_1}},
\end{equation*}
at that $C = 64A^3f_2\varsigma$. The Hubble parameter
\begin{equation}
H_{ln}=\frac{A(2\cosh(At)^2-3)}{3\cosh(At)\sinh(At)}
-\frac{C_1\cosh(At)^2}{6Af_1\tilde{D}_2((\tilde{D}_2\tanh(At)^2+\tilde{D}_0)
\tanh(At)}.
\end{equation}
corresponds to this solution.

\subsection{Exponential Potential}

In paper~\cite{CN} the exponential potential has been consider in addition
to the cubic potential and approximate solutions for equation
(\ref{equagen}) have been obtained.

Let $V(\phi)=C_1e^{\alpha\phi}$. At $f_0=0$ and $C=0$ solutions for
(\ref{eomN}) are elementary functions and have the following form:
\begin{equation}\label{EXPsol}
\phi_{exp}(t)=\frac{1}{\alpha}\ln\left(\frac{4f_1}{C_1\alpha^2(t-t_0)^2}\right),
\qquad H_{exp}(t)=\frac{1}{t-t_0}.
\end{equation}
Note that the obtained Hubble parameter $H_{exp}(t)$ is proportional to
the Hubble parameter, which has been used in paper~\cite{CN}, and to the Hubble
parameter, obtained in the case of cubic potential (formula
(\ref{H3})).

\subsection{Power Potential}

Let us consider solutions in the case of the potential
$V(\phi)=C_1\phi^n$. At $f_0=0$ equation (\ref{Localeomgenpot}) is as
follows:
\begin{equation}
{\dot\phi}^2=\frac{C_1^2f_2n^2(n-1)\phi^n-C_1Cf_2n(n-1)\phi-Cf_1^2\phi^{3-n}}{f_1f_2C_1n(n-1)(n-2)}
\end{equation}

At $C=0$ this equation is equivalent to
\begin{equation}
{\dot\phi}^2=\frac{C_1n\phi^n}{f_1(n-2)}
\end{equation}
and has a solution in the form of the elementary function:
\begin{equation}
\phi_n(t)=
2^{2/(n-2)}\left(\frac{f_1}{C_1n(n-2)(t-t_0)^2}\right)^{1/(n-2)}.
\end{equation}
The corresponding Hubble parameter is equal to
\begin{equation}
H_n(t)=\frac{3n-4}{3(n-2)(t-t_0)}.
\end{equation}

At $n=4/3$ we get a particular solution to equation~\eqref{equagen} in the Minkowski space:
\begin{equation}
\phi_m(t)={}\pm \frac{2C_1\sqrt{-2 f_1C_1}}{27f_1^2}(t-t_0)^3.
\end{equation}
Note that in the Minkowski space exact bounded at all values of $t$
solutions for nonlocal equations with power potentials have been found
in paper~\cite{AJ}.

\section{COSMOLOGICAL MODEL WITH A NONLOCAL \\ SCALAR FIELD AND A $k$-ESSENCE FIELD}

The goal of this Section is to show, that the supplement of the $k$-essence scalar field~$\Psi$  allows
to obtain a system of the Einstein equations, which has an exact solution, at that the Hubble parameter
 and the nonlocal field are given by formulae~\eqref{Hphi} and~\eqref{Localeomgenpot} respectively. The
$k$-essence models are considered in cosmology both as inflation models~\cite{Mukhanov1,GostCond,Creminelli},
and as dark energy models~\cite{Mukhanov2,Rendall,KVW,BMV,CreminelliDE},
(see also~\cite{VikmanPhD} and references therein).

Let us consider the following action
\begin{equation}
S_2=\int\,d^4x\,\sqrt{-g}
\biggl(\frac{R}{16\pi G_{\mathrm N}}+\frac{1}{g_\text{o}^2}
\biggl(\frac{1}{2}\phi\mathcal F(\Box_g)\phi
-V(\phi)\biggr)-\mathcal P(\Psi,X)\biggr),
\label{39}
\end{equation}
where $X\equiv-g^{\mu\nu}\partial_\mu\Psi\partial_\nu\Psi$.
In the Friedmann--Robertson--Walker metric the function~$\Psi$ depends only on time, so $X=\dot\Psi^2$.

Following paper~\cite{CreminelliDE}, we select the pressure in the following form
\begin{equation}
\mathcal P(\Psi,X)
=\frac{1}{2}(p_q(\Psi)-\varrho_q(\Psi))
+\frac{1}{2}(p_q(\Psi)+\varrho_q(\Psi))X
+\frac{1}{2}M^4(\Psi)(X-1)^2.
\label{40}
\end{equation}

We consider functions $p_q(\Psi)$, $\varrho_q(\Psi)$, and $M^4(\Psi)$ as arbitrary differentiable functions. The energy density of the
$k$-essence field is equal to
\begin{equation}
\mathcal E(\Psi,X)
=(p_q(\Psi)+\varrho_q(\Psi))X
+2M^4(\Psi)(X^2-X)-\mathcal P(\Psi,X).
\label{41}
\end{equation}

The Einstein equations, which have been obtained using the variation of $S_2$,
have the following form:
\begin{equation}
\begin{aligned}
&3H^2=8\pi G_{\mathrm N}(\varrho+\mathcal E),
\\
&2\dot H+3H^2=-8\pi G_{\mathrm N}(p+\mathcal P).
\end{aligned}
\label{42}
\end{equation}

Varying the action $S_2$, we also get equation~\eqref{equagen} and equation for the $k$-essence scalar field~$\Psi$
\begin{equation}
\dot{\mathcal E}={}-3H(\mathcal E+\mathcal P).
\label{43}
\end{equation}

In the Friedmann--Robertson--Walker metric  the energy--momentum tensor~\eqref{TEV} has the following form
\begin{equation}
T_{\mu\nu}=g_{\mu\nu}\mathrm{diag}(-\varrho,p,p,p),
\label{44}
\end{equation}
where
\begin{align*}
&\varrho=\frac1{g_\text{o}^2}
\biggl(\sum_{n=1}^\infty\frac{f_n}{2}
\biggl(\sum_{l=0}^{n-1}\partial_t\Box^l\phi
\partial_t\Box^{n-1-l}\phi
+\sum_{l=1}^{n-1}\Box^l\phi\Box^{n-l}\phi\biggr)
-\frac{f_0}{2}\phi^2+V(\phi)\biggr),
\\
&p=\frac1{g_\text{o}^2}\biggl(\sum_{n=1}^\infty
\frac{f_n}{2}\biggl(\sum_{l=0}^{n-1}
\partial_t\Box^l \phi\partial_t\Box^{n-1-l}\phi
-\sum_{l=1}^{n-1}\Box^l\phi\Box^{n-l}\phi\biggr)
+\frac{f_0}{2}\phi^2-V(\phi)\biggr).
\end{align*}

Let $\phi_2$ is a solution to system~\eqref{eomN}
at $N=2$. Using  $\Box_g^2\phi_2=C/f_2$,
we get
\begin{equation}
\varrho(\phi_2)=E_{00}(\phi_2)+W(\phi_2),\qquad
p(\phi_2)=E_{00}(\phi_2)-W(\phi_2),
\label{45}
\end{equation}
where
\begin{align*}
&E_{00}(\phi_2)=\frac{1}{2g_\text{o}^2}\left(f_1(\partial_t\phi_2)^2
+2f_2\partial_t\phi_2\partial_t\Box_g\phi_2
+f_3(\partial_t\Box_g\phi_2)^2\right),
\\
&W(\phi_2)=\frac1{g_\text{o}^2}
\biggl(\frac{f_2}{2}(\Box_g\phi_2)^2
+\frac{f_3C}{f_2}\Box_g\phi_2
+\frac{f_4C^2}{2f_2^2}
-\frac{f_0}{2}\phi_2^2+V(\phi_2)\biggr).
\end{align*}

The $k$-essence models (without additional fields) have one useful property.
For any real differentiable function~$H_0(t)$ there exist such differentiable functions
$\varrho_q(\Psi)$ and $p_q(\Psi)$, that functions $H_0(t-t_0)$ and $\Psi(t)=t-t_0$ are a particular solution
to system~\eqref{42}--\eqref{43}.
This property can be generalized on the case of the models with an additional nonlocal scalar field, which are described by action~\eqref{39}.
Indeed, at $\Psi(t)=t-t_0$ we obtain
\begin{equation}
\mathcal E=\varrho_q(\Psi)=\varrho_q(t-t_0),\qquad
\mathcal P=p_q(\Psi)=p_q(t-t_0).
\label{46}
\end{equation}
Substituting into~\eqref{42} expression of~$\varrho_q$ and $p_q$, we get
\begin{equation}
\begin{aligned}
&\varrho_q(\Psi)=\varrho_q(t-t_0)
=\frac{3}{8\pi G_{\mathrm N}}H^2(t-t_0)-\varrho(t-t_0),
\\
&p_q(\Psi)=p_q(t-t_0)
=-\varrho_q(t-t_0)-\varrho(t-t_0)-p(t-t_0)
-\frac{1}{4\pi G_{\mathrm N}}\dot H(t-t_0).
\end{aligned}
\label{47}
\end{equation}
It is easy to see that system~\eqref{42}--\eqref{43}
and equation~\eqref{equagen} have the exact particular solution, at that functions
$H(t-t_0)$ and $\phi(t-t_0)$ are the obtained solution of equation~\eqref{equagen}
and $\Psi(t)=t-t_0$.

So, the algorithm  to obtain exact solutions is as follows: for the given potential~$V(\phi)$ we find $H(t)$ and
$\phi(t)$, calculate the energy--momentum tensor and substitute the obtained values into~\eqref{47}.
The obtained values of~$\varrho_q$ and $p_q$ give the exact solvable model with a nonlocal scalar field and the
 $k$-essence field. The function $M(\Psi)$ can be selected arbitrarily.

Let us illustrate this scheme on the simplest example, connected with a cubic potential, and
consider solution~\eqref{elemsol1}--\eqref{H3}.
Conditions~\eqref{24} of existence of this solution leave constants~$B_3$
and~$B_2$ arbitrary. Using the arbitrariness of $B_2$, we can, without loss of generality, put
$f_0=0$. Also, for compactness of the record
we put $t_0=0$. For solution~\eqref{elemsol1}--\eqref{H3} we get
\begin{equation}
\Box_g\phi_1=\frac{16f_1}{3B_3t^4},
\label{48}
\end{equation}
therefore,
\begin{equation}
E_{00}
=\frac{32f_1^2(f_1t^4+16f_2t^2+64f_3)}
{9B_3^2t^{10}},\qquad
W=\frac{128f_2f_1^2}{9g_\text{o}^2B_3^2t^8}
+\frac{1}{g_\text{o}^2}V(\phi_1).
\label{49}
\end{equation}
Consequently we get
\begin{align*}
&\varrho_q(\Psi)
=\frac{B_2^3-27B_0B_3^2}{27g_\text{o}^2B_3^2}
+\frac{25}{24\pi G_{\mathrm N}\Psi^2}
-\frac{160f_1^3}{27g_\text{o}^2B_3^2\Psi^6}
-\frac{640f_2f_1^2}{9g_\text{o}^2B_3^2\Psi^8}
-\frac{2048f_3f_1^2}{9g_\text{o}^2B_3^2\Psi^{10}},
\\
&p_q(\Psi)
=-\frac{B_2^3-27B_0B_3^2}{27g_\text{o}^2B_3^2}
-\frac{5}{8\pi G_{\mathrm N}\Psi^2}
-\frac{32f_1^3}{27g_\text{o}^2B_3^2\Psi^6}
-\frac{128f_2f_1^2}{3g_\text{o}^2B_3^2\Psi^8}
-\frac{2048f_3f_1^2}{9g_\text{o}^2B_3^2\Psi^{10}}.
\end{align*}

Thus, we can make a conclusion that the adding of the $k$-essence field  allows
to construct a self-consistent system of the Einstein equations for the obtained solution to equation~\eqref{equagen}.
The choice of the $k$-essence scalar field  as an additional field is reasoned by the goal to get  a self-consistent system
 of the Einstein equations with no restriction on the type of the potential and solutions.

\section{CONCLUSIONS}

In this paper the method of the search of exact solutions to the nonlocal
field equation~\eqref{equagen} in the Friedmann--Robertson--Walker metric has been proposed. It has been demonstrated
that for an arbitrary potential besides of linear or quadratic ones there
exists a particular two-parameter solution, which can be found in
quadratures.

The string field theory inspired case of a cubic potential has been consider in detail.
It is shown that solutions are either Weierstrass elliptic functions or
degenerate elliptic functions, for example,  solutions in terms of hyperbolic
tangent have been obtained. Exact solutions in terms of elementary functions have been found  for logarithmic, exponential
and power potentials as well.

Note that the proposed method allows to find a solution only for the
field equation, but not for the whole system of the Einstein equations. It has been shown that
the supplement of the $k$-essence scalar field allows
to construct a self-consistent system of the Einstein equations for the obtained solution to equation~\eqref{equagen}. The
localization of the Einstein equations in cosmological models
with a nonlocal scalar field and an arbitrary potential is an actual problem, which studying
requires distinct investigations\footnote{The localization method has been constructed only in the case of linear or quadratic potential~\cite{Vernov2010}.}.

\subsection*{Acknowledgements}
The author is grateful to the organizers of the International
Bogolyubov confe\-rence ''Problems of Theoretical and Mathematical
Physics'' (Moscow--Dubna, Russia, August 20--26, 2009) for the possibility to
present results of my work and financial support. I wish to express my
thanks to I.Ya.~Aref'eva and N.~Nunes for useful and stimulating
discussions. This research is supported in part by RFBR grant
08-01-00798, grant of Russian Ministry of Education and Science
NSh-4142.2010.2, and by Federal Agency for Science and Innovation under
state contracts 02.740.11.5057 and 02.740.11.0244.

\end{document}